# Quantum phases in f-Electron Systems

Karan Singh[1], K. Mukherjee[2] and A. M. Jayannavar[3]

[1]Institute of Physics, Sachivalaya Marg, Bhubaneswar 751005, India

[2]School of Basic Sciences, Indian Institute of Technology Mandi, Mandi 175005, Himachal Pradesh, India

[3]D.M.S Mandal's Bhaurao Kakatkar College, Belgaum 590001, Karnataka, India

**Abstract**

Quantum fluctuations and related phase transitions are of current interest from the viewpoint of fundamental physics and technological applications. Quantum phase implies a region where the quantum fluctuations of energy scale $\hbar\omega$ dominates over the thermal energy $k_BT$. Presence of quantum phase leads to unconventional and unexpected physical phenomena like Kondo effect, non-Fermi liquids, ordered magnetic state, and Fermi liquids, etc. In this framework, Ce-based metallic compounds, exhibiting correlated electron phenomena, emerged as prototypical systems to study the various quantum phases. In these systems considerable efforts have been made, both experimentally and theoretically, to overcome the problems related to the comprehensive understanding of correlated quantum phases. In this article, various aspects related to quantum phases in $CeNiGe_2$, CeGe and CeAlGe are summarized, mainly focusing on the structural and physical properties.

## 1. Introduction

In simple metals, valence electrons move freely throughout the lattice and many properties are satisfactorily explained by the classical model given by Drude without invoking electron-electron interactions. For example, resistivity decreases linearly with decreasing temperature due to reduced scattering of electrons by lattice vibrations with some residual resistivity, owing to impurities at zero Kelvin. However, when the electron-electron interactions are slowly turned on, the non-interacting states smoothly and continuously evolve into interacting state and low-energy excitation of a system of mobile fermions is described in term of well-defined fermionic particles called "quasiparticle" [1]. Quasiparticle carries charge (-e), spin (±1/2) and a mass $m^*$ (the mass being renormalized by interactions) and it can be described by good quantum numbers, like $p$ (momentum) and $\sigma$ (spin projection) which in a translation invariant system are occasionally supplemented by additional internal quantum number, such as energy band index etc. Quasiparticles obey a dispersion relation $\omega = \varepsilon_{p\sigma}$ ($\varepsilon$ is energy) and it allows one to make use of the powerful methods of the weak coupling or mean-field-like theories. The life-time of the quasiparticle is long-lived and it behaves like Fermi liquids at low temperature. The resistivity of Fermi liquids varies quadratically ($\rho \sim T^2$), while heat capacity varies linear ($C \sim \gamma T$), where $\gamma$ is the summerfeld coefficient. The value of the $\gamma$ determines the amount of correlation in a system and is proportional to effective mass [2]. In the case of strongly correlated electron systems, Ce-based intermetallics have been extensively investigated, where strong coupling exists between the 4*f* localized moments and the conduction band [3]. This coupling results the formation of two energy scales: i) Ruderman–Kittel–Kasuya–Yosida (RKKY) interactions and ii) Kondo interactions. The RKKY interaction tends to order magnetic moments of the localized 4*f*-shell. On the other hand, the Kondo effect tends to screen localized magnetic moments due to polarization of conduction electrons. Under the application of external parameters, such as magnetic field, doping etc, there is competition between RKKY and Kondo interactions, resulting in various novel states; e.g. antiferromagnetic ordering, non-Fermi liquid, quantum critical point, unconventional superconductivity, etc [4-7].

In this article, several novel states existing in $CeNiGe_2$, CeGe, and CeAlGe, are presented. The 4f level of $Ce^{3+}$ ions lies close to the Fermi level, resulting in strong hybridization between the 4f localized electrons and the conducting d- and s-electrons. The outline of this article is as follow: in the beginning, the crystal structures of $CeNiGe_2$, CeGe, and CeAlGe are discussed. Later, the magnetization, electrical transport and heat capacity behavior of these compounds along with the phenomena such as RKKY interaction, metallic behavior, Kondo effect, Sommerfeld coefficient, non-Fermi liquid and quantum criticality are discussed.

## 2. Crystal structure
Crystal structure of $CeNiGe_2$ and CeGe are orthorhombic, with space groups *Cmcm* and *pnma* respectively (figure 1a and 1b). The orthorhombic $CeNiGe_2$ has lattice parameters a = 4.25Å, b = 16.79Å, c = 4.21Å with atomic positions as Ce (0.0, 0.105, 0.25), Ni (0.0, 0.32, 0.25), Ge1 (0.0, 0.46, 0.25), and Ge2 (0.0, 0.74, 0.25). The Ge1 square sheets and Ge2 zigzag chains are bridged by Ni atoms into a 3D open framework. In this structure, it is believed that magnetic contribution of Ni is almost negligible in comparison to Ce ions [8]. In orthorhombic CeGe, Ce atoms per unit cell are four. It has lattice parameters as a = 8.358Å, b = 4.082Å, c = 6.028Å and atomic positions as Ce (0.178, 0.25, 0.615), Ge (0.039, 0.25, 0.135). CeAlGe crystallizes in tetragonal structure (with space group $I4_1md$), which is a non-centrosymmetric structure (figure 1c). The lattice parameters are a = 4.28Å, b = 4.28Å, c = 14.69Å, and the atomic positions are as Ce (0.5, 0.5, 0.57), Al (0.0, 0.0, 0.66), Ge1 (0.0, 0.0, 0.48).

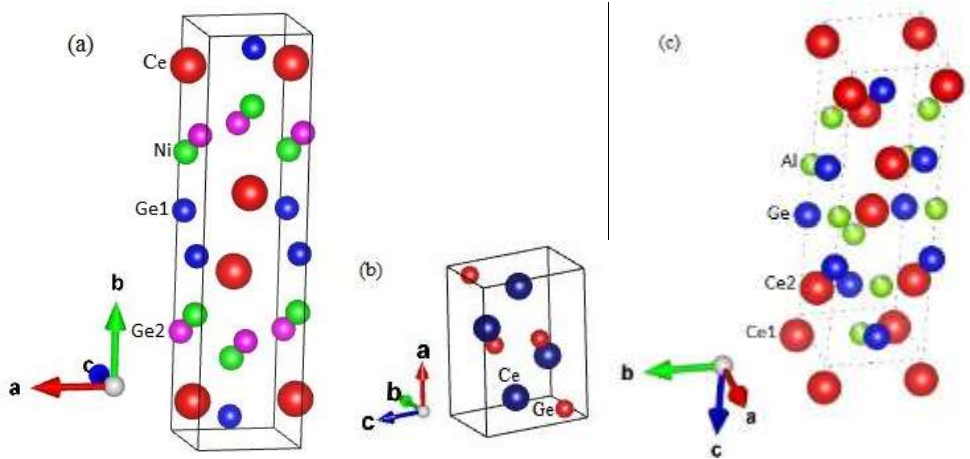

Figure 1: Crystal structure for (a) $CeNiGe_2$ (b) CeGe (c) CeAlGe

## 3. Magnetism

In metallic Ce-based compounds, the magnetism is due to RKKY interactions, where localized moments interact through mobile electrons of the conduction band. These interactions are defined in term of Hamiltonian, such as: H = J(R) $S_i.S_j$ (where $S_i$ and $S_j$ are the localized magnetic moment at site i and j, respectively and J(R) is the exchange interaction term, which varies in an oscillatory behavior with distance R, between the magnetic moments). RKKY interaction depends on the conduction electron density at the Fermi level $g(E_F)$ ($E_F$ = Fermi energy) and temperature of RKKY interaction is defined as [9]

$T_{RKKY} \sim J^2 g(E_F)$ …….. (1)

The RKKY interaction strength dominates over the thermal energy at lower temperatures, resulting in magnetic ordering temperature. For examples, the $CeNiGe_2$, CeGe and CeAlGe compounds has the magnetic transition temperatures at 4.1 K, 10.7 K, and 5.2 K, respectively (figure 2a, 2b, and 2c), due to presence of RKKY interactions.

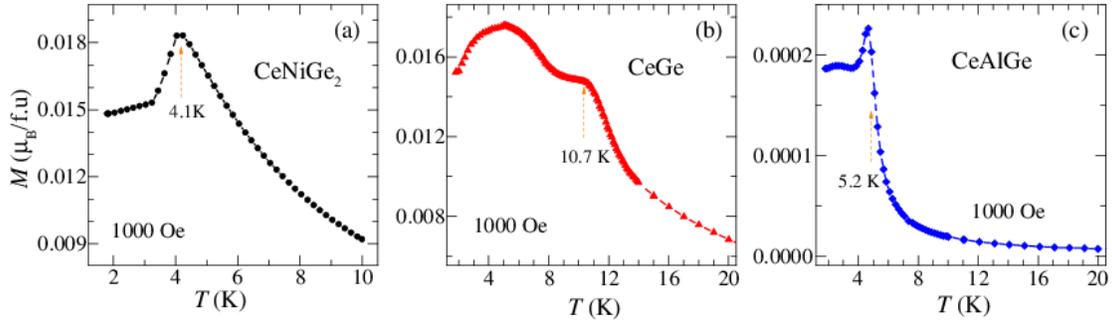

Figure 2: Temperature dependence magnetization (a) CeNiGe$_2$ (b) CeGe (c) CeAlGe. These data are taken from the Refs [8, 10, 11]. Copyright by the Elsevier, Nature and Taylor & Francis publishing group for figure (a), (b) and (c), respectively.

In all compounds, the magnetic susceptibility ($\chi$) in the paramagnetic region (above the transition temperature) can be defined by the Curie Weiss law:

$\chi(=M/H) = C/(T-\theta)$ ….. (2)

where $C$ is the Curie constant and $\theta$ is the Curie-Weiss temperature. $\theta$ is positive for ferromagnetic and negative for antiferromagnetic interactions. All compounds CeNiGe$_2$, CeGe and CeAlGe show negative Curie-Weiss temperature, indicating the dominance of antiferromagnetic type interactions [8, 10, 11]. The Curie constant $C$ is equal to $\mu_0 N_A \mu_{eff}^2 \mu_B^2 / 3k_B$, from where the effective moment $\mu_{eff}$ can be estimated. The obtained effective moments of these compouds is around 2.6±0.2 $\mu_B$, which is close to the effective moments of free ion of $Ce^{3+}$. Below transition temperature, magetization increases with increasing fields. The magnetization value is the highest for CeAlGe and the smallest for CeGe at 70000 Oe as seen from $M$ vs $H$ plot (figure 3).

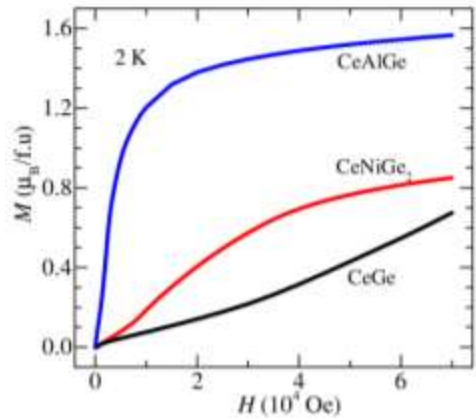

Figure 3: Field dependence magnetization at 2 K for CeNiGe$_2$, CeGe and CeAlGe. These data are taken from the Refs [8, 10, 11]. Copyright by the Elsevier, Nature and Taylor & Francis publishing group for data of CeNiGe$_2$, CeGe and CeAlGe, respectively.

The difference in the magnetization could be due to crystal electric field effect of the Ce-ion, due to which the degeneracy of the multiplet ground state is lifted into three doublet states, resulting in dipolar magnetic moments. However, in some systems, quartet state is formed and accounts for higher order moments; e.g. quadrupolar, octupolar moments [4, 10, 12]. Higher-order terms can be extracted from the non-linear susceptibility as:

$M/H = \chi_1 + \chi_3 H^2 + ....$(3)

where $\chi_1$ is the dipolar susceptibility and $\chi_3$ is associated with the higher-order susceptibility (as quadrupolar). Out of the three compounds, the CeNiGe$_2$ and CeGe show the higher-order susceptibility [figure 4a, 4b], which has experimental extracted using protocol given in ref [12]. It is responsible for the complex magnetic state at low temperatures for CeNiGe$_2$ and CeGe [10, 12].

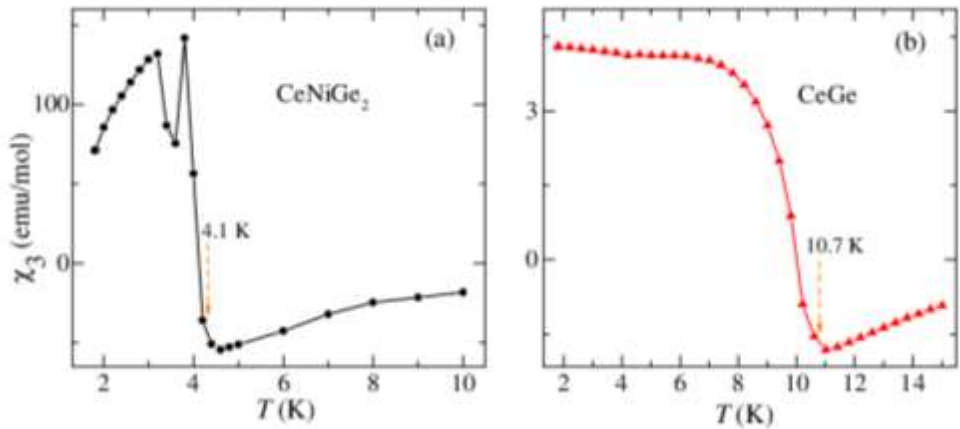

Figure 4: Temperature dependece $\chi_3$ for (a) CeNiGe$_2$ and (b) CeGe. These data are taken from the Refs [10, 12]. Copyright by the Taylor & Francis and Nature publishing group for figure (a) and (b), respectively.

## 4. Electrical transport

The electrical transport through materials is a broad and complex field, however, in this section; the basic scattering aspect related to the simple metal, Kondo effect, and Fermi/non-Fermi liquids is covered. In metal, the valence electrons moves randomly throughout space and net conductivity is zero. When external voltage is applied, the electric field is generated which drift the electrons from one end to another, resulting in conductivity in metal. Here, it is assumed that electron-electron interaction is negligible, but the scattering of the electron by the lattice vibration impede the electron path and resistivity ($\rho$) develops, as explained by Ohm law. This resistivity is temperature dependent, which decreases with decreasing temperature. It is due to the fact that lattice vibration decreases with decreasing temperature and thus reduces scattering. In addition, for metals residual resistivity can exist at very low temperature or at zero temperature, where the lattice vibration have almost died out, due to scattering of the electrons with defects, impurities and vacancies. The presence of a magnetic impurity in the sea of conduction electrons can result in logarithmic increase in resistivity (log ($T$)) with decreasing temperature; eg. Fe in Cu metal [13, 14]. This phenomenon is known as the Kondo effect. It is the unusual scattering of the conduction electrons, and its mechanism describes the many-body scattering process subject to low energy quantum mechanical degree of freedom. Out of three compounds discussed here, only $CeNiGe_2$ show a Kondo type signature with Kondo temperature ($T_K$), around 13 K (figure 5a, highlighted by squre shaded region). It can be understood by the virtue of Heisenberg's uncertaninty principle; the localized electrons spontaneously tunnel to the conduction band at Fermi level, if timescale limited by the uncertaninty principle, an electron from the conduction band tunnels to the available localized state. In this process, the spins of electrons are exchanegd and render an intermediate state at the Fermi level. This intermediate state effectivelly scatters electrons near the Fermi level resulting in resistivity varies with log($T$). In other compounds (CeGe and CeAlGe), the screening effect is expected to be weak. Hence, the signature of Kondo temperature is not visible. Figure 5a, 5b and 5c depicts temperature dependent of resistivity for $CeNiGe_2$, CeGe and CeAlGe respectively. In addition to the Kondo temperature a discontinuity appear at the transition temprature (as marked by arrow in the figure 5 plot). Below the magnetic transition, the resistivity decreases due to magnetic ordering due to reduced scattering of conduction electrons.

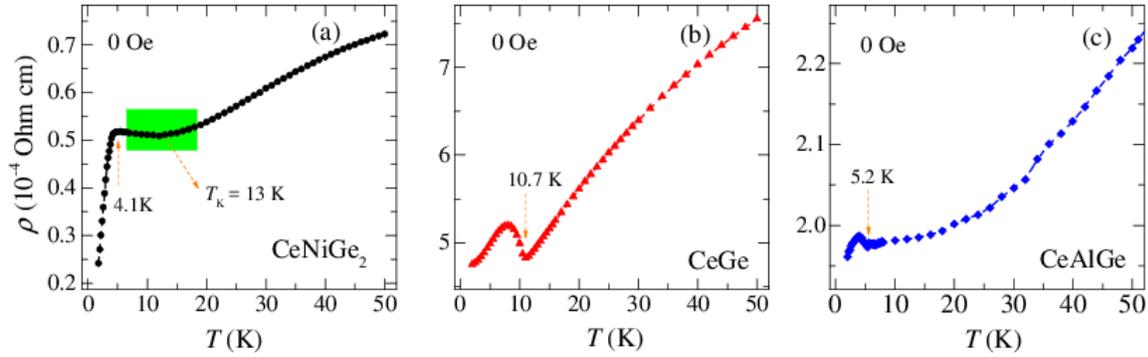

Figure 5: Temperature dependence of resistivity (a) $CeNiGe_2$ (b) CeGe (c) CeAlGe These data are taken from the Refs [8, 10, 11]. Copyright by the Elsevier, Nature and Taylor & Francis publishing group for data of $CeNiGe_2$, CeGe and CeAlGe, respectively.

At low temperatures, Fermi liquid may be expected. However, the $T^2$ variation of resistivity has not been followed which violates the Fermi liquid phenomenon. Fermi liquid provides the basis for understanding metals in terms of weakly interacting electrons. In the case of strongly interacting electrons, the non-Fermi liquid phenomena is suspected due to suppressing magnetic state with doping non-magnetic ion; e.g. in $Ce_{0.24}La_{0.76}Ge$ [15]. In CeGe, non-magnetic La subsitutition at the Ce-site suppresses the magnetic ordering and shows the non-Fermi liquid behavior for $Ce_{0.24}La_{0.76}Ge$ [15]. The resistivity varies linearly with $T^{1.6}$ dependence (below 3.5 K) (figure 6a). The non-Fermi liquid behavior is caused due to disordered 4f spins induced by the La ion. In CeGe, the temperature of the magnetic transition is about 10.7 K, below which the magnetic interaction energy is larger than the Kondo interaction energy. La substituion adds disorder in magnetic state, suppressing the magnetic transition. This results in reduced magnetic energy which becomes approximately comparable to Kondo energy. On this energy scale, non-Fermi liquid could be expected as seen in $Ce_{0.24}La_{0.76}Ge$. Disorder driven non-Fermi liquid can be verifed by the dynamical mean field theory of the spin glass quantum critical point, which follows the resistivity scaling of $\rho(T, H) - \rho(0, H) \propto t^{3/2} \Psi(t/\Delta)$ (figure 6b), where $H$ is magnetic field, $t = T/T_0$ ($T_0$ is the temperature as a function of $J$) and $\Psi(x)$ is the scaling function [15]. $\Delta$ is defined as: $\Delta = \Delta_0 + 2(\Delta_0)^{1/2} t \{[1 + t/2(3)^{1/2}\Delta_0]^{1/2} - 1\}$, where $\Delta_0 = r + (H/H_0)^2$; $r (= 1-J/J_c)$ and $H_0 = J_c/(g\mu_B)$, where $g$ (gyromagnetic ratio) = 2 for Ce ion and $\mu_B$ is the Bohr magnetron.

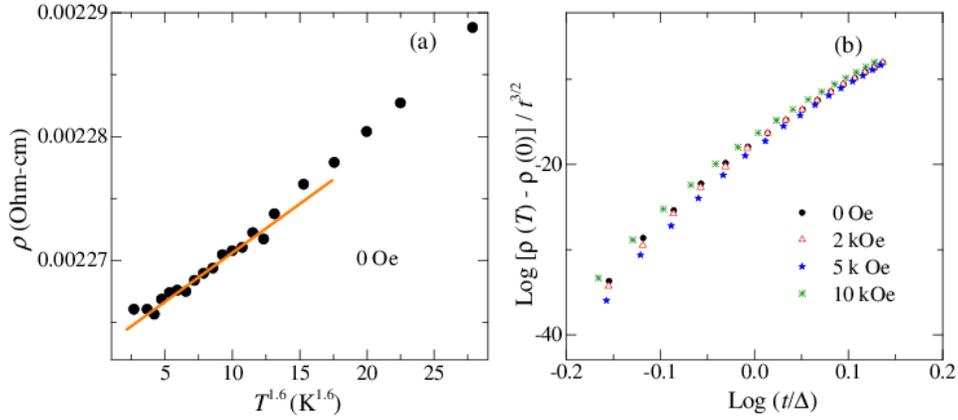

Figure 6: (a) $T^{1.6}$ dependent resistivity at 0 Tesla for $Ce_{0.24}La_{0.76}Ge$. (b) Scaling plots of the resistivity in low temperature regime and field range 0-10 kOe for same compound. These data are taken from the Ref [15]. Copyright by the IOP publisshing group.

In metals with strong electron correlations, non-Fermi liquid is often found close to the quantum critical point (QCP). It occurs when the phase transition temperature of a system is suppressed to absolute zero and a quantum superposition of order and disorder develops. Around QCP, unusual power law in the temperature dependent resistivity occur ($\rho \sim T^x$, $x < 2$) and heat capacity ($C$) diverges logrithmally ($\sim -\log(T)$), or power law ($T^\alpha$, $\alpha$ is temperature exponent) [16]. The mysterious quantum fluctuations are considered to be responsible for the profound transformation of physical properties close to QCP.

## 5. Heat capacity

Heat capacity is one of the most essential thermodynammic properties of metals. It is defined as the amount of heat required to raise the temperature of a given quantity of matter by one degree. For metal, the heat capacity can be written as a sum of the electron and phonon contribution below the Debye and the Fermi temperature as:

$C = \gamma T + \beta T^3$ ……..(4)

where $\gamma$ is the Sommerfeld coefficient associated with electronic contribution and $\beta$ is coeffecient associated with phonon contribuiton, which can be determined experimentally. At high temperatures, the lattice vibration is very large ($T^3$ dependence). With cooling the lattice vibration vanishes faster, resulting in heat capacity that varies almost linearly due to electronic contribution at low temperature. The value of the $\gamma$ determines the strength of the electron-electron correlations. For metal with negligible electron-electron electrons, the γ is around 0.695 mJmol$^{-1}$K$^{-2}$ for Cu. When electron-electron interaction turned on, it is found that γ increases with increasing interactions. In strongly correlated electron systems; eg. heavy fermions, γ is found to be one thousand times more than copper metal. Experimentally, the γ is determined from the slope of equation (4) above magnetic transition. For CeNiGe$_2$, CeGe, and CeAlGe, the values of γ are 433 mJ/mol K$^2$, 267 mJ/mol-K$^2$, and 22 mJ mol$^{-1}$ K$^{-2}$, respectively [8, 10, 11]. These values indicate large electron – electron correlations. Among the three compounds, it is noted that CeNiGe$_2$ has the largest value, due to strongly correlated electron behavior, found mostly in heavy fermion compounds. The heat capacity also shows the magnetic transition at the same temperature where it is seen in the magnetization and resitivity curves for all compounds, due to magnetic ordering (figure 7a, 7b, and 7c).

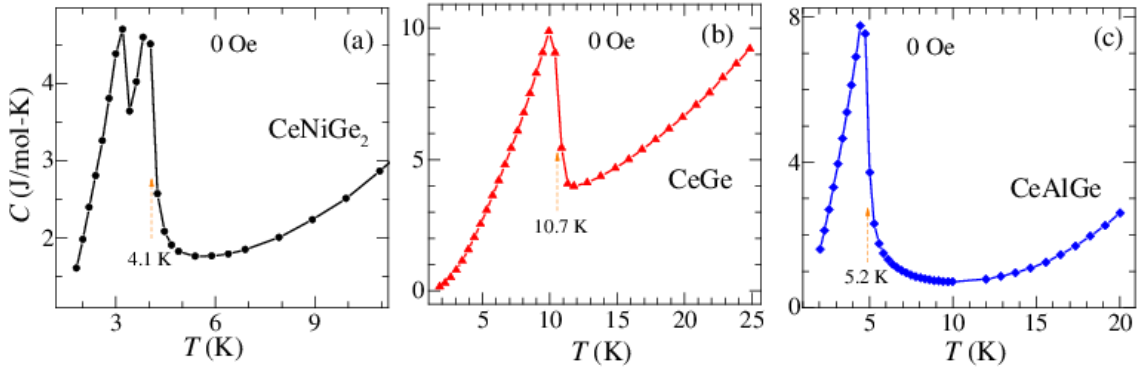

Figure 7: Temperatre dependence of heat capacity (a) CeNiGe$_2$ (b) CeGe (c) CeAlGe. These data are taken from the Refs [8, 10, 11]. Copyright by the Elsevier, Nature and Taylor & Francis publishing group for data of CeNiGe$_2$, CeGe and CeAlGe, respectively.

Further, heat capacity is an interesting tool to investigate the quantum phase transition close to the QCP. The quantum fluctutions of the energy scale $\hbar\omega$ is enhanced to the QCP, where ω is the characteristic frequency of the quantum oscillations and is inversely proportional to the correlation time. With external parameter, eg. magnetic field ($H$), the correlation time ($\xi_\tau$) and correlation length (ξ) can be described as [17]: $\xi_\tau \sim \xi^z$, $\xi \sim |r|^{-\nu}$ ($r = (H-H_C)/H_C$, $H_C$ = critical field) and ν is the exponent of correlation length, $z$ is the dynamical critical exponent, depending on the dynamic of the order parameter. At critical field, the heat capacity vary with $T^n$ (inset of figure 8), where $n = d/z - 1$ is the temperature exponent, $d$ is dimensionality. Heat capacity ($C'$) (by subtracting the lattice contribution) at $H_C$ and at finite $T$ can be defined as [17]:

$$\Delta C'/T^{d/z} = \Psi (H/T^{\beta\delta/\nu z}) \quad \ldots\ldots\ldots (5)$$

where $\Delta C'/T^{d/z} = C'(T, H) - C'(T, H_C)$, $d/z$ and $\beta\delta/\nu z$ are scaling factor coupled with the parameters, magnetic field and temperature and $\Psi(H/T^{\beta\delta/\nu z})$ is scaling function. The scaling of heat capacity indicates to presence of quantum criticality. In compound $Ce_{0.6}Y_{0.4}NiGe_2$, such heat capacity scaling has been observed at critical field $H_C \sim 50$ kOe (figure 8), indicating to quantum criticality in this compound [17]

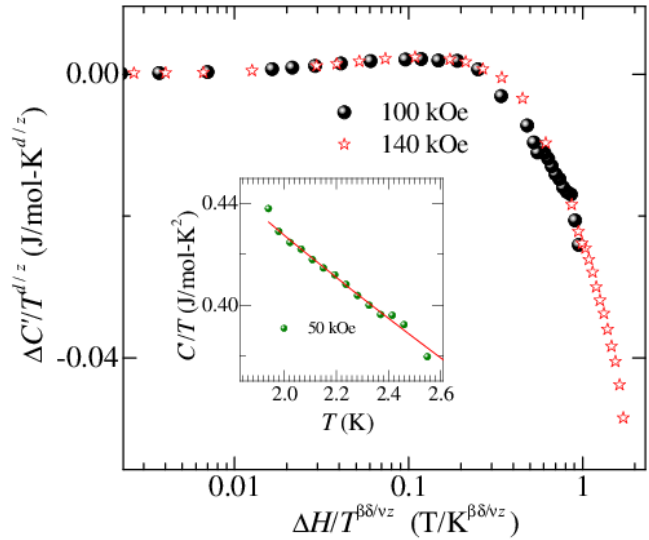

Figure 8: Scaling of field dependence heat capacity for $Ce_{0.6}Y_{0.4}NiGe_2$ compound. Inset: temperature dependence of $C'/T$ at field at 50 kOe. Solid red lines are the fit of equation $C'/T \sim T^n$. These data are taken from the Ref [17]. Copyright by the Elsevier publisshing group.

## 6. Conclusions

In this article, an overview of the basic physical propeprties of $CeNiGe_2$, $CeGe$, and $CeAlGe$ is presented. These compounds are promising candidates for the finding of various quantum phases associated with magnetic orderings, multipolar moments, Kondo effect, non-Fermi liquid, and quantum criticality. Using techniques like magnetization, electrical transport and heat capacity, it is realized that localized 4f electrons and there interactions with conducting d- and s- electrons are the key ingredients for various exotic quantum phase transitions shown by these compounds. Despite developing these exotic phases, it is believed that the subject of quantum phases of correlated electrons is at a very early stage; eg. the theoretical understanding of non-Fermi liquid is still controversial. Likely, in systems of this types, unconventional superconductivity has also been observed, about which it is believed that this may open avenues for understanding high temperature superconductivity.

**Acknowledgments**

KS thanks D. Topwal for informative discussions and inputs. A.M.J. thanks DST, India, for financial support through J.C. Bose Fellowship